\documentclass[
,draft            
  ]  {aipproc}

\layoutstyle{6x9}
\usepackage{amsfonts}
\usepackage{amssymb}
\def\lp{\stackrel{\leftarrow}{\partial}}
\def\rp{\stackrel{\rightarrow}{\partial}}
\def\be{\begin{eqnarray}}
\def\ee{\end{eqnarray}}
\def\*{\star}

\begin{document}

\title{Deformation Quantization of Nambu Mechanics}

\author{Cosmas K Zachos}{  address={
High Energy Physics Division,
Argonne National Laboratory, Argonne, IL 60439-4815, USA 
{\sl zachos@hep.anl.gov} }
}

\author{Thomas L Curtright}{
  address={
Department of Physics, University of Miami,
Box 248046, Coral Gables, Florida 33124, USA   
{\sl curtright@physics.miami.edu} } 
}

\begin{abstract} 
Phase Space is the framework best suited for quantizing superintegrable 
systems---systems with more conserved quantities than 
degrees of freedom. In this quantization method, the symmetry algebras 
of the hamiltonian invariants are preserved most naturally, as illustrated
on nonlinear $\sigma$-models, 
specifically for Chiral Models and de Sitter 
$N$-spheres. Classically, the dynamics of superintegrable models such as these 
is automatically also described by Nambu Brackets involving the extra symmetry
invariants of them. The phase-space quantization worked out 
then leads to the quantization of the corresponding Nambu Brackets, 
validating Nambu's original 
proposal, despite excessive fears of inconsistency which have arisen over 
the years. This is a pedagogical talk based on \cite{sphere,CQNB}, 
stressing points of interpretation and care needed in appreciating the 
consistency 
of Quantum Nambu Brackets  in phase space.  For a parallel discussion in 
Hilbert space, see T Curtright's contribution in these Proceedings,
[hep-th/0303088].
\end{abstract}
\maketitle

\section {Introduction}
Highly symmetric quantum systems are often integrable, and, in 
special cases, superintegrable and exactly solvable \cite{winter}.
A superintegrable system of $N$ degrees of freedom has more than
$N$ independent invariants, and a maximally superintegrable one has $2N-1$ 
invariants. The classical evolution of all functions in phase space for 
such systems is alternatively specified through 
Nambu Brackets (NB) \cite{nambu,hietnambu,takhtajan,nutku}.
 However, quantization of NBs has been considered 
problematic ever since their inception. We find that it need not be.

In the case of velocity-dependent potentials, when quantization 
of a classical system presents operator ordering ambiguities 
involving $x$ and $p$, the general consensus has long been 
\cite{velo,lakshmanan, higgs, leemon} to select those orderings
in the quantum hamiltonian which maximally preserve the symmetries 
present in the corresponding classical hamiltonian. Even for simple systems, 
such as $\sigma$-models considered here, 
such constructions may become involved and needlessly technical.

There is a quantization procedure ideally suited to this problem of selecting 
the quantum hamiltonian which maximally preserves integrability.
In contrast to conventional operator 
quantization, this problem  is addressed most cogently in Moyal's 
phase-space quantization formulation  \cite{moyal,cfz}, reviewed in 
\cite{czreview}. 
The reason is that the variables involved in it 
(``phase-space kernels" or ``Weyl-Wigner inverse transforms of operators") 
are c-number 
functions, like those of the classical phase-space theory, 
and have the same interpretation, although they involve $\hbar$-corrections
(``deformations"), in general---so $\hbar\rightarrow 0$ reduces to the 
classical expression.  It is only the detailed algebraic structure 
of their respective brackets and composition rules which contrast with 
those for the variables of the classical theory. This complete formulation is 
based on the 
Wigner Function (WF), which is a quasi-probability distribution function in 
phase-space, and comprises the kernel function of the density matrix.
Observables and transition amplitudes are  phase-space integrals of kernel
functions weighted by the WF, in analogy to statistical mechanics. Kernel
functions, however, unlike ordinary classical functions, multiply through the
$\*$-product, a noncommutative, associative, pseudodifferential operation, 
which encodes the entire quantum mechanical action and whose antisymmetrization 
(commutator) is the Moyal Bracket (MB) \cite{moyal,cfz,czreview}, the
quantum analog of the Poisson Bracket (PB). 

Groenewold's correspondence principle theorem \cite{groenewold} 
(to which van Hove's extension is routinely attached \cite{vanHove}) 
evinces that there is no invertible linear map from {\em all} 
functions of phase space $f(x,p), g(x,p),...,$ to hermitean operators in 
Hilbert space ${\mathfrak Q}(f)$, ${\mathfrak Q}(g),...,$ such that
the PB structure is preserved consistently, 
\be 
{\mathfrak Q} (\{ f,g\})= \frac{1}{i \hbar} ~
[ ~  {\mathfrak Q}(f)  , {\mathfrak Q} (g)~\Bigr ] ~.
\ee
This cannot be achieved, in general.

Instead, the Weyl correspondence map \cite{weyl,czreview} from functions to
ordered operators,
\be
{\mathfrak W}(f) \equiv \frac{1}{(2\pi)^2}\int d\tau d\sigma 
dx dp ~f(x,p) \exp (i\tau ({ {\mathfrak  p}}-p)+i\sigma ({ {\mathfrak  x}}-x)),
\ee
specifies an associative (and non-commutative) $\*$-product 
\cite{groenewold,czreview}, 
\be
\star \equiv \exp \left ( {i\hbar \over 2} 
(\lp_x \rp_{p} - \lp_{p} \rp_x )\right ),
\ee
so that $
{\mathfrak W} (f\star g) =
{\mathfrak W} (f) ~   {\mathfrak W} (g) $,
and thus 
\be 
{\mathfrak W} (\{\! \{ f,g \}\!  \} )= \frac{1}{i \hbar} ~
\Bigl [ {\mathfrak W}(f)  , {\mathfrak W} (g) \Bigr ] ~.
\ee
Here, the Moyal Bracket is defined as 
\be
\{\!\{ f, g \}\!\} \equiv {f \* g - g\* f \over i\hbar} ~,
\ee
and as $\hbar\rightarrow 0$, ~~MB $\rightarrow$ PB. That is,
it is the MB instead of the PB which maps invertibly to the quantum commutator,
completing Dirac's original proposal.
This relation underlies the foundation of phase-space quantization 
\cite{groenewold,czreview}. 

Conversely, given an arbitrary operator 
${\mathfrak F} ({\mathfrak x}, {\mathfrak p})$ 
consisting of operators ${\mathfrak x}$ and 
${\mathfrak p}$, one might imagine rearranging it by use of Heisenberg 
commutations to a canonical completely symmetrized 
(Weyl-ordered) form, in general with $O(\hbar )$ terms generated in the 
process. It might then be mapped uniquely to its Weyl-correspondent c-number 
kernel function $f$ in phase space ${\mathfrak x}  \mapsto x$,  and 
${\mathfrak p} \mapsto p$,  ~
${\mathfrak W}^{-1}( {\mathfrak F}) = f(x,p,\hbar)$. (In practice, there is 
a more direct inverse transform formula \cite{groenewold,czreview} which 
bypasses a need to rearrange to a canonical Weyl ordered form explicitly.)
Clearly, operators differing from each other by different orderings of 
their ${\mathfrak x}$s and ${\mathfrak p}$s correspond to kernel 
functions $f$ coinciding with each other at $O(\hbar^0)$, but different 
in $O(\hbar)$, in general.      Thus, a survey of all 
alternate operator orderings in a problem with such ambiguities amounts,
in deformation quantization, to a survey of the ``quantum correction" 
$O(\hbar)$ pieces of the respective kernel functions, ie the inverse 
Weyl transforms of those operators, and their study is greatly systematized 
and expedited. Choice-of-ordering problems then reduce to purely
$\*$-product algebraic ones, as the resulting preferred orderings are 
specified through particular deformations in the c-number kernel expressions 
resulting from the particular solution in phase space. 

Hietarinta \cite{hietarinta} has investigated in this phase-space quantization
language the simplest integrable systems of velocity-dependent potentials. In
each system, he has promoted the vanishing of the Poisson Bracket (PB) of the
(one) classical invariant $I$ (conserved integral)  with the hamiltonian, 
$\{ H,I\}=0$, to the vanishing of its (quantum) Moyal Bracket (MB) with the
hamiltonian, $\{\! \{ H_{qm} ,I_{qm} \}\!  \} =0$. 
This dictates quantum corrections, addressed
 perturbatively in $\hbar$, specifically $O(\hbar^2)$ corrections to the $I$s
and $H$ ($V$), needed for quantum symmetry. The expressions found are quite
simple, as the systems chosen are such that the polynomial character of the
$p$s, or suitable balanced combinations of $p$s and $q$s, ensure collapse or
subleading termination of the MBs. The specification of the symmetric
hamiltonian then is complete, since, as indicated,
 {\em the quantum hamiltonian in terms of
classical phase-space variables corresponds uniquely to the Weyl-ordered
expression for these variables in operator language}. 

Here, nonlinear $\sigma$-models (with explicit illustrations 
on $N$-spheres and chiral models)  are utilized to argue for the general 
principles of power and convenience in isometry-preserving quantization in
phase space, for large numbers of invariants, in principle (as many as the
isometries of the relevant manifold). In the cases illustrated, the number of
algebraically independent invariants matches or exceeds the dimension of the
manifold, leading to superintegrability \cite{winter}, whose impact is best 
surveyed through Classical Nambu Brackets (CNBs). 
The procedure of determining the proper
symmetric quantum Hamiltonian then yields remarkably compact and elegant
expressions.

Briefly, we find that the symmetry generator invariants are undeformed by 
quantization, but the Casimir invariants  of their MB algebras {\em are} 
deformed, in accord with the Groenewold-van Hove theorem.
Hence, the hamiltonians are also deformed by terms $O(\hbar^2)$, as they 
consist of quadratic Casimir invariants. Their spectra are then read off 
through group theory, suitably adapted to phase space \cite{sphere}. 
The basic principles are illustrated for the simplest 
curved manifold, the two-sphere, and are then 
generalized to larger classes of symmetric manifolds such as 
Chiral Models and $N$-spheres.

Given the elegant quantization of maximally superintegrable systems in phase 
space, as worked out here, and further given their classical dynamics 
specified by CNBs, one might wonder {\em why} quantization of 
NBs had been deemed to be beset with inconsistencies. Direct 
comparison of Nambu's early quantization  prescription \cite{nambu} 
to the conventional quantum answers now at hand actually reveals that 
there are {\em no} insurmountable inconsistencies. Comparison to the standard
Moyal deformation quantization vindicates Nambu's quantization  prescription
which yields identical results, and thus also identical to standard Hilbert
space quantization, for all systems discussed here. We find that, in 
exceptional cases, the
algebra of invariants ensures that the Quantum Nambu Brackets (QNBs) are
derivations. However, in the more generic case, we find that, 
even though QNBs are not derivations, they still involve derivations 
entwined in symmetrized (Jordan)
products with invariants, and are fully consistent.

\section {spheres and chiral models} 
Consider a particle on a curved manifold, in integrable one-dimensional 
$\sigma$-models, 
\begin{equation}
L(q,\dot{q})={1\over2} g_{ab}(q)~\dot{q}^a\dot{q}^b,
\label{Posaunenstoss}
\end{equation}
so that 
\begin{equation}
p_a={\partial L\over{\partial \dot{q}^a}}= g_{ab}~ \dot{q}^b,
\qquad  \dot{q}^a= g^{ab} p_b~.
\end{equation}
Thus, 
\begin{equation}
H(p,q)={1\over2} g^{ab} p_a p_b  \qquad   (=L).
\label{Amsel}
\end{equation}
The isometries of the manifold generate the conserved integrals of the 
motion \cite{bcz}. The classical equations of motion are 
\be
\dot{p}_a= -\frac{g^{bc}_{,a}}{2}~ p_b p_c = \frac{g_{bc,a}}{2}~ \dot{q}^b
 \dot{q}^c.
\ee

As the simplest possible nontrivial illustration, consider a particle on a 
2-sphere of unit radius, $S^2$.  In Cartesian coordinates (after the 
elimination of $z$, so $q^1=x, q^2=y$), one has, for $a,b=1,2$:
\begin{equation} 
g_{ab}= \delta_{ab}+ {q^a q^b \over u}, \qquad  
g^{ab}= \delta_{ab}- q^a q^b , \qquad  \det g_{ab} =\frac{1}{u},
\qquad u\equiv 1-x^2-y^2~.
\end{equation}
($u$ is the sine-squared of the latitude, since this represents 
the orthogonal projection of the globe on its equatorial plane).
The momenta are then
\be
p_a= \dot{q}^a + q^a {h\over u}  
= \dot{q}^a + q^a (q\cdot p)  ~, \qquad   
h\equiv -\dot{u}/2 = x \dot{x} + y \dot{y} ~. 
\ee
The classical equations of motion here amount to 
\be
\dot{p}_a = p_a ~q\cdot p ~, \qquad \hbox{i.e.}  \qquad   
\ddot{q}^{~a}+ q^a \left ( {\dot{h}\over u} + {h^2 \over u^2}\right ) =0.
\ee

It is then easy to find the three classical invariants, the 
components of the conserved angular momentum in this nonlinear realization,
\be
L_z= x p_y - y p_x ~, \qquad L_y= \sqrt {u} ~ p_x  ~, \qquad 
L_x=  -\sqrt {u}~ p_y ~.
\ee
(The last two are the de Sitter ``momenta", or nonlinearly realized 
``axial charges" corresponding to the ``pions" $x,y$ of the $\sigma$-model:
linear momenta are, of course, not conserved).
Their PBs close into $so(3)$, 
\be
\{L_x,L_y \}= L_z ~,\qquad   
 \{L_y,L_z \}= L_x ~,\qquad   
\{L_z,L_x\}= L_y~. \label{Hacivat} 
\ee
Thus, it follows algebraically that their PBs with the Casimir invariant 
${\bf L}\cdot{\bf L}$ vanish. Naturally, since $H= {\bf L}\cdot{\bf L}/2$, 
they are manifested to be time-invariant, 
\be 
\dot{{\bf L}}=\{ {\bf L}, H \}=0.
\ee

In quantizing this system, operator ordering 
issues arise, given the effective velocity(momentum)-dependent potential.
In phase-space quantization, one may insert $\*$-products 
 in strategic points and orderings of the variables of (\ref{Amsel}), 
to maintain integrability. That is, the classical invariance 
expressions (PB commutativity),
\be
\{ I, H \}=0,
\ee
are to be promoted to quantum invariances (MB commutativity),
\be
\{\!\{ I_{qm}, H_{qm}\}\!\} \equiv {I_{qm}\* H_{qm}- H_{qm}\* 
I_{qm}\over i\hbar} =0~.
\ee
Here, this argues for a (c-number kernel function) hamiltonian of the form
\be
H_{qm}= \frac{1}{2} \left ( L_x\*  L_x +   
L_y\*  L_y +   L_z \*  L_z \right ) .   \label{siornionios}
\ee
The reason is that, in this realization, the algebra (\ref{Hacivat}) is 
promoted to the corresponding MB expression {\em without any modification}, 
since all of its MBs collapse to PBs by the linearity in momenta of the 
arguments: all corrections $O(\hbar)$ vanish. 
Consequently, these particular invariants are undeformed by quantization, 
${\bf L}= {\bf L}_{qm}$. 
As a result, given associativity for $\*$, the corresponding quantum  
quadratic Casimir invariant
${\bf L}\cdot\* {\bf L}$ has vanishing MBs with ${\bf L}$ (but not vanishing 
PBs\footnote{Likewise, in comportance with the Groenewold-van Hove theorem, 
the above classical hamiltonian $H$ does not MB-commute with the invariants.
}), and automatically serves as a symmetry-preserving hamiltonian.
The specification of the maximally symmetric quantum hamiltonian is thus 
complete.
 
The $\*$-product in this hamiltonian trivially evaluates to yield the quantum 
correction to (\ref{Amsel}),
\be
H_{qm} = H + \frac{\hbar^2}{8} \Bigl ( \det g -3 \Bigr ). \label{okanumaihi}
\ee
One might wish to contrast this quantum correction to the free-space 
angular-momentum quantum correction for zero curvature in the underlying 
manifold familiar from atomic physics \cite{dahlPecs}: that one too 
is $O(\hbar^2)$,
but it is a constant, reflecting the vanishing curvature. 
Predictably, on the North pole above, $u=1$, and these expressions 
coincide.
This difference and pole coincidence carries over for all dimensions, 
as evident in the quantum correction (\ref{sN}) for $S^N$ derived below.

In phase-space quantization \cite{moyal,cfz,czreview}, the WF (the kernel 
function of the density matrix) evolves according to Moyal's 
equation \cite{moyal}, 
\be 
{\partial f \over \partial t} 
=\{\! \{H_{qm} , f \}\! \} ~. \label{mukuru}
\ee
But, in addition, for pure stationary states, the spectrum is specified 
by the fundamental $\*$-genvalue equations \cite{dbf,cfz} satisfied by 
the respective WFs, 
\be 
H_{qm}  (x,p)\star f(x,p) &=& f(x,p)   \star H_{qm} (x,p)
 \nonumber \\
&=&H_{qm} \left (x+{i\hbar\over 2} \rp_p ~,~ 
p-{i\hbar\over 2} \rp_x\right ) ~ f(x,p)= E ~f(x,p)~.
\label{NordhauserSchattensaft}
\ee

The spectrum of this specific hamiltonian (\ref{siornionios}), then, is 
proportional to the $\hbar^2 l(l+1)$ 
spectrum of the $SO(3)$  Casimir invariant ${\bf L}\cdot\* {\bf L}=
L_{+}\* L_{-}+ L_z\* L_z - \hbar L_z$, for integer $l$ \cite{bffls}. It can 
be produced algebraically by the standard recursive ladder operations 
in $\*$ space which obtain in the operator formalism Fock space \cite{sphere}.
Similar $\*$-ladder arguments and inequalities apply directly in phase space 
to all Lie algebras.

The treatment of the 3-sphere $S^3$ is very similar, with some
significant differences, since it also accords to the standard chiral model 
technology. The metric and eqns of motion, etc, 
are identical in form to those above, except now
$u\equiv 1-x^2-y^2-z^2= 1/\det g $,   
$~h\equiv -\dot{u}/2 = x \dot{x} + y \dot{y}+z \dot{z}$,
and $a,b=1,2,3$. However, the description simplifies upon utilization
of Vielbeine, $g_{ab}= \delta_{ij} V_a^i V_b^j$ and 
$g^{ab} V_a^i V_b^j=\delta^{ij}$.

Specifically, the Dreibeine, are either left-invariant, or 
right invariant \cite{cuz}: 
\be
^{(\pm)}V_a^i=   \epsilon^{iab} q^b \pm  \sqrt{u}~ g_{ai} ~,
\qquad 
^{(\pm)}V^{ai}=   \epsilon^{iab} q^b \pm  \sqrt{u}~ \delta^{ai} .
\ee
The corresponding right and left conserved charges (left- and right-invariant,
respectively) then are
\be
R^i= ~^{(+)}V^i_a  ~ \dot{q}^a  = ~^{(+)}V^{ai} p_a ~, \qquad 
L^i=~^{(-)}V^i_a   ~ \dot{q}^a =~^{(-)}V^{ai} p_a ~.
\ee
More intuitive than those for $S^2$  are the linear combinations into Axial and 
Isospin charges (again linear in the momenta),
\be
{{\bf R}- {\bf L}\over 2}= \sqrt{u} ~ {\bf p}\equiv {\bf A} ,  \qquad
{{\bf R}+{\bf L}\over 2}= {\bf q}\times {\bf p}  \equiv {\bf I}.
\ee

It can be seen that the ${\bf L}$s and the ${\bf R}$s have 
PBs closing into standard $su(2)\otimes su(2)$, ie,  
$su(2)$ relations within each set, and vanishing 
between the two sets. Thus they are seen to be constant, since the hamiltonian 
(and the lagrangian) can, in fact, be written in terms of either 
quadratic Casimir invariant,
\be
H= \frac{1}{2}  {\bf L}\cdot {\bf L}=
  \frac{1}{2}  {\bf R}\cdot {\bf R} =L  ~.     \label{specS3}
\ee

Quantization consistent with integrability thus proceeds as 
above for the 2-sphere, since the MB algebra collapses to PBs again,
and so the quantum invariants {\bf L} and {\bf R} again coincide with 
the classical ones, without deformation (quantum corrections). The 
$\*$-product is now the obvious 
generalization to 6-dimensional phase-space. The eigenvalues of the relevant
Casimir invariant are now $j(j+1)$, for half-integer $j$. 
However, this being a chiral model ($G\otimes G$), the symmetric
quantum hamiltonian is simpler that the previous one, since it can  now
also be written geometrically as 
\be
H_{qm}= \frac{1}{2} (p_a V^{ai}) \* ( V^{bi}  p_b) =
\frac{1}{2}\left (g^{ab} p_a p_b + {\hbar^2 \over 4} 
\partial_a V^{bi} \partial_b V^{ai} \right )  . \label{quantS3} 
\ee
The Dreibeine throughout this formula can be either $~^{+}V^i_a$ or
$~^{-}V^i_a$, corresponding to either the right, or the 
left-acting quadratic Casimir invariant.   
The quantum correction then amounts to 
\be
H_{qm}- H = {\hbar^2 \over 8} \Bigl ( \det g  - 7 \Bigr ) .  \label{shozaburo}
\ee

In general, the above discussion also applies to all chiral models,
with $G\otimes G$ replacing $su(2)\otimes su(2)$ above. 
Ie, the Vielbein-momenta 
combinations $V^{aj}p_a$ represent algebra generator invariants, 
whose quadratic Casimir 
group invariants yield the respective hamiltonians, and whence the properly
$\*$-ordered quantum hamiltonians as above. (We follow the conventions of 
\cite{bcz}, taking the generators of $G$ in the defining representation to
be $T_j$.) 

That is to say, for 
\be
i U^{-1} \frac{d}{dt}   U=~^{(+)}V^j_a T_j \dot{q}^a =~^{(+)}V^{aj} p_a T_j  ~,
\qquad i U\frac{d}{dt}   U^{-1} =~^{(-)}V^{aj} p_a T_j ~,
 \ee
it follows that the PBs of the left- and right-invariant charges 
$~^{(\pm)}V^{aj} p_a =\frac{i}{2}Tr T_{j}U^{\mp 1}\frac{d}{dt}U^{\pm 1} $ 
close to the identical Lie algebras,
\be
\{ ^{(\pm)} V^{aj}p_a ,   ^{(\pm)}V^{bk}p_b \} = -2 f^{jkn}~^{(\pm)}V^{an}p_a ~,
\label{Lie}
\ee
and PB commute with each other, 
\be
\{ ~^{(+)} V^{aj}p_a ,  ^{(-)}V^{bk}p_b \}=0. \label{karagoz} 
\ee
These two statements are proven directly in \cite{sphere}.

MBs collapse to PBs by linearity in momenta as before, and the hamiltonian is 
identical in form to (\ref{quantS3}). The quantum 
correction in  (\ref{quantS3}) to amounts to 
\be
H_{qm}- H = {\hbar^2 \over 8}\Bigl (\Gamma^b_{ac}~ g^{cd} \Gamma^a_{bd} -
f_{ijk}f_{ijk}\Bigr ) ,  
\ee
(reducing to (\ref{shozaburo}) for $S^3$). 
The spectra are given by 
the Casimir eigenvalues for the relevant 
algebras and representations.

For the generic sphere models, $S^N$, the maximally symmetric hamiltonians 
are the quadratic Casimir invariants of $so(N+1)$, 
\be
H=\frac{1}{2}  P_a P_a + \frac{1}{4} L_{ab} L_{ab} ~,  
\ee
where 
\be
P_{a}=\sqrt{u} ~p_a~, \qquad \qquad L_{ab}=q^a p_{b}-q^{b} p_a~, 
\ee
for $a=1,\cdots,N$, 
the de Sitter momenta and angular momenta of $so(N+1)/so(N)$.
All of these $N (N+1)/2$ sphere translations and rotations are symmetries of 
the classical hamiltonian. 
  
Quantization proceeds as in $S^2$,  maintaining conservation of all $P_{a}$ and
$L_{ab}$,
\be
H_{qm}=\frac{1}{2}  P_a \* P_a + \frac{1}{4} L_{ab} \* L_{ab} ~, 
\label{Moosbrugger}
\ee
and hence the quantum correction is 
\be
H_{qm}- H = {\hbar^2 \over 8}\left ( \frac{1}{u}- 1 -N(N-1) \right) .\label{sN}
\ee
The spectra are proportional to the 
Casimir eigenvalues $l(l+N-1)$ for integer $l$ \cite{bffls}. For $N=3$ 
of the previous section, 
this form is reconciled with the Casimir expression for (\ref{specS3}) as
$l=2j$, and agrees with \cite{lakshmanan,higgs,leemon}.

\section {Maximal Superintegrability and the Nambu Bracket} 

All the models considered above have extra invariants 
beyond the number of conserved quantities in involution
(mutually commuting) required for integrability in the Liouville sense. 
The most systematic way of accounting for such additional invariants, 
and placing them all on a more equal footing, even when they do not all 
simultaneously commute, is the NB formalism.  
 
For example, the classical mechanics of a particle on an N-sphere
as discussed above may be summarized elegantly through Nambu Mechanics in 
phase space \cite{nambu,takhtajan}. Specifically, \cite{hietnambu,nutku},
 in an $N$-dimensional space, and thus $2N$-dimensional 
phase space, motion is confined on the constant surfaces specified by 
the algebraically independent integrals of the motion (eg, $L_x,L_y,L_z$ 
for $S^2$ above.) Consequently, the phase-space velocity 
${\bf v}=(\dot{{\bf q}},\dot{{\bf p}})$  is always perpendicular to the 
$2N$-dimensional phase-space gradients $\nabla=(\partial_{\bf q}, 
\partial_{\bf p})$ of all these integrals of the motion, and 
$\nabla\cdot {\bf v}=0$.

As a consequence, if there are $2N-1$ algebraically independent
such integrals $L_i$, possibly including the hamiltonian, 
(ie, the system is maximally superintegrable \cite{winter}), 
the phase-space velocity
must be proportional \cite{hietnambu} to the cross-product of all those 
gradients, and hence the motion is fully specified for any phase-space 
function $k({\bf q},{\bf p})$
 by a phase-space Jacobian which amounts to the Nambu Bracket:
\be 
{dk\over dt}&=&\nabla k \cdot {\bf v}  ~\propto ~ 
\partial_{i_1} k ~ \epsilon^{i_1i_2...i_{2N}} ~ 
\partial_{i_2} L_{i_1} ... \partial_{i_{2N}}L_{2N-1} \nonumber   \\
& = & {\partial (k,L_1,...,...,L_{2N-1}) \over \partial 
(q_1,p_{1},  q_2,p_{2},  ...,q_{N},p_{N})   } 
\equiv  \{k,L_1,...,L_{2N-1}\},\label{muhacir} 
\ee 
as an alternative to Hamiltonian mechanics.
For instance, for the above $S^2$, 
\be
{dk\over dt}= {\partial (k,L_x,L_y ,L_z ) \over \partial (x,p_x,y,p_y)} 
=\left\{  k,L_x \right\}\left\{ L_y,L_z \right\}-\left\{  k,L_y \right\}  
\left\{ L_x  ,L_z \right\}-\left\{  k,L_z \right\}\left\{ L_y,L_x \right\}~.
\label{Rum}
\ee
For the more general $S^N$, one now has a choice of $2N-1$ of the 
$N(N+1)/2$ invariants of $so(N+1)$; one of several possible expressions is 
\be
\frac{dk}{dt}= { \left(  -1\right)  ^{\left(  N-1 \right)}  \over 
 P_{2}P_{3}\cdots P_{N-1} }
\frac{\partial\left(  k,P_{1},L_{12},P_{2},L_{23},P_{3},\cdots,P_{N-1}, 
L_{N-1\;N},P_{N}\right)  }{\partial\left(  x_{1},p_{1},x_{2},p_{2},
\cdots,x_{N},p_{N}\right)  } ~, \label{omunene}
\ee
where $P_{a}=\sqrt{u} ~ p_a$, for $a=1,\cdots,N$, 
 and $L_{a,a+1}=q^a p_{a+1}-q^{a+1} p_a$, for $~a=1,\cdots,N-1$. 

In general \cite{Sahoo,takhtajan}, NBs, being Jacobian determinants, 
possess all antisymmetries of such; being linear in all derivatives, they 
also obey the Leibniz rule of partial differentiation, 
\be
\{ k(L,M) ,f_1,f_2,... \} = 
\frac{\partial k}{\partial L} \{ L ,f_1,f_2,... \}   +
\frac{\partial k}{\partial M} \{ M ,f_1,f_2,... \} . \label{Leibniz}
\ee
Thus, an entry in the NB algebraically dependent on  the remaining entries 
leads to a 
vanishing bracket.  For example, it is seen directly from above that
the hamiltonian is constant, 
\be
{dH\over dt}=\left\{ { {\bf L}\cdot{\bf L} \over 2}, ...\right \}=0,
\ee
since each term of this NB vanishes. Naturally, this also applies to
all explicit examples discussed here, as they are all maximally 
superintegrable.

Finally, the impossibility to antisymmetrize 
more than $2N$ indices in $2N$-dimensional phase space, 
\be
\epsilon ^{ab....c [ i } \epsilon^{j_1 j_2 ...j_{2N}  ]   } =0,
\ee
leads to the so-called Fundamental Identity, \cite{Sahoo,takhtajan},  
slightly generalized here \cite{sphere,CQNB}, 
\be
&\phantom{=}&\{ V \{A_1,...,A_{m-1},A_m \} , A_{m+1},...,A_{2m-1}   \} \nonumber \\
&+&\{A_m, V\{A_1,...,A_{m-1},A_{m+1}\},A_{m+2},...,A_{2m-1}\}\nonumber \\
&+&...+ \{A_m,...,A_{2m-2}, V\{A_1,...,A_{m-1},A_{2m-1} \} \} \nonumber \\
&=&\{ A_1,...,A_{m-1},V\{ A_m , A_{m+1},...,A_{2m-1} \}   \}. \label{FI}  
\ee
Its name is inapposite, however, to the extent that it merely reflects the fact 
that CNBs, being linear in derivatives, obey Leibniz's rule (they are 
derivations \cite{Sahoo,CQNB}); even though it has sometimes been 
analogized to the Jacobi Identity, unlike the Jacobi identity, it is not 
fully antisymmetrized in all of its arguments,  nor does it reflect 
associativity in the abstract \cite{CQNB,Hanlon,Azcarraga}.

The proportionality function $V$ in (\ref{muhacir}),
\be
{dk\over dt}= V \{k,L_1,...,L_{2N-1} \}, \label{tuzsuz} 
\ee
has to be a time-invariant \cite{nutku} if it has no explicit time 
dependence. This is seen from consistency of (\ref{tuzsuz}),
application of which to 
\be
{d \over dt} \Bigl ( V\{A_1,...,A_{2N} \}\Bigr )
= \dot{V}\{A_1,...,A_{2N}  \} + V\{\dot{A_1},...,A_{2N} \} +...+
V\{A_1,...,\dot{ A}_{2N} \},
\ee
yields  
\be
V\{V\{A_1,...,A_{2N} \},L_1,...,L_{2N-1}\} &=& \dot{V}\{A_1,...,A_{2N} \} \\
+V\{V \{A_1,L_1,...,L_{2N-1} \} ,...,A_{2N} \} &+&...+
V\{A_1,..., V \{A_{2N} ,L_1,...,L_{2N-1} \}\}, \nonumber
\ee
and, by virtue of (\ref{FI}), $\dot{V}=0$ follows.

Actually, PBs result from a maximal reduction of NBs, 
by inserting $2N-2$ phase-space coordinates and summing over them, thereby
taking {\em symplectic traces}, 
\be
\left\{  L,M\right\}  = \frac{1}{\left(  N-1\right)!}
\left\{  L,M,x_{i_{1}},
p_{i_{1}},\cdots,x_{i_{N-1}},p_{i_{N-1}}\right\},
\ee 
where summation over all $N-1$ pairs of repeated indices is 
understood.
Fewer traces lead to relations between NBs of maximal rank, $2N$, and those
of lesser rank, $2k$,
\be
\left\{  L_{1},\cdots,L_{2k}\right\} = \frac{1}{\left(  N-k\right)!}
\left\{
L_{1},\cdots,L_{2k},x_{i_{1}},p_{i_{1}},\cdots,x_{i_{N-k}},p_{i_{N-k}%
}\right\}
\ee 
(which is one way to define the lower rank NBs for $k\neq1$), or
between two lesser rank NBs \cite{CQNB}. Note that 
$\left\{  L_{1},\cdots,L_{2k}\right\}  $ 
acts like a Dirac Bracket (DB) up to a normalization, 
$\{  L_{1} ,L_{2}\}  _{DB}$, where the fixed additional entries 
$L_{3},\cdots,L_{2k}$ in the
NB play the role of the constraints in the DB \cite{bracketpamd}. 
(In effect, this has been previously observed, eg, \cite{nutku}, 
for the extreme case $N=k$, without symplectic traces.)

As a simplest illustration, consider $N=k=2$ for the system (\ref{Rum}),
but now taking $L_x,L_y$ as second-class constraints:
\be
\{f,g,L_x,L_y\}= \{f,g\}  \{L_x,L_y \}+\{f,L_x\}  \{L_y,g\}- 
\{f,L_y\}  \{L_x,g \}\equiv  \{ L_x ,L_y\} ~\{f,g\}_{DB}.
\ee
That is, 
\be
\{f,g\}_{DB}= \Bigl (\{L_x,L_y\}\Bigr ) ^{-1}~   \{f,g,L_x,L_y\},  
\ee
so that, from (\ref{FI}) with $V= (\{L_x,L_y\})^{-1}=1/L_z$ (also see 
\cite{nutku}), it follows that the 
Dirac Brackets satisfy the Jacobi identity,
\be
\{ \{f,g\}_{DB},h\}_{DB}+
\{ \{g,h\}_{DB},f\}_{DB}+
\{ \{h,f\}_{DB},g\}_{DB}=0~,
\ee
a property usually established by explicit calculation \cite{bracketpamd}, in 
contrast to this derivation. Naturally, $\{f,L_x,L_y,L_z\}=\{f,H \}_{DB}$.

\section {Nambu Quantization}
The quantization method of NBs considered by Nambu in an operator context 
\cite{nambu}, when applied to the phase-space formalism, motivates 
defining QNBs as fully antisymmetrized associative $\*$-product sequences,
\be 
\left [ A, B \right ]_{\*}   &\equiv &  
A\* B-B\* A= i\hbar~ \{\!\{A,B \}\!\}, \nonumber \\
\left [ A,B,C \right ] _{\*} &\equiv & A \star B\* C-A \* C \star B
+B\star C\star A-B\star A\star C+C\star A\star B-C\star B\star A ,\nonumber \\
\left[ A,B,C,D\right] _{\star } &\equiv &A\star \left[ B,C,D\right]
_{\star }-B\star \left[ C,D,A\right] _{\star }+C\star \left[
D,A,B\right] _{\star }-D\star \left[ A,B,C\right] _{\star }  \nonumber \\
&=&  [A,B]_\star \star [C,D]_\star + [A,C]_\star \star [D,B]_\star 
+ [A,D]_\star \star [B,C]_\star \nonumber \\ 
&\phantom{a}& +[C,D]_\star \star [A,B]_\star + [D,B]_\star \star [A,C]_\star 
+ [B,C]_\star \star [A,D]_\star~,   \label{QNB}
\ee 
etc. These antisymmetrized $\*$-products are used in the quantum theory 
instead of the previous jacobians. As for the above 4-QNB, the resolution of 
all even-QNBs into symmetrized $\*$-products of 2-brackets (MBs) 
reflecting the full antisymmetry of the structure is a general 
useful result \cite{CQNB}.

These QNBs, consisting of associative strings of $\*$-products, 
satisfy Generalized Jacobi Identities \cite{Hanlon,Azcarraga,CQNB},
but not,  in general \cite{dfst}, the Leibniz property (\ref{Leibniz}) and the 
consequent Fundamental Identity (\ref{FI}). 
The loss of the latter two properties is a 
subjective shortcoming, contingent on the specific application context. 
Objectively, this approach is in
agreement with the $\*$-product quantization of all the examples given above,
the N-spheres, Chiral Models, and, in addition, $n$-dimensional isotropic
oscillator systems \cite{CQNB}.

Specifically, for $S^2$, it follows directly from (\ref{QNB}) and 
(\ref{Hacivat})  (with MBs supplanting PBs, $\{\!\{ L_x,L_y \}\! \} = L_z$,
$\{\!\{  L_y,L_z \}\! \}= L_x $,
$\{\!\{ L_z,L_x\}\! \}= L_y$) that the Moyal Bracket with the hamiltonian 
(\ref{siornionios}) equals Nambu's QNB, for an arbitrary function $k$ of phase 
space,
\be 
\left[ k,L_x ,L_y ,L_z \right] _{\*} 
=  i \hbar ~ \left[ ~k, {\bf L}\cdot\* {\bf L} ~\right]_{\star } 
= -2 \hbar^2 ~  \{\! \{ k, H_{qm} \}\!\} ~,
\ee 
so that for $k$ with no explicit time dependence, 
\be
\frac{dk}{dt}   =\frac{-1}{2\hbar ^{2}}\Bigl [ k,L_{x},L_{y},L_{z}\Bigl ]_{\*}.
\ee
This provides a good quantization for (\ref{tuzsuz}), for this 
particular system (\ref{Rum}).
For $\hbar\rightarrow 0$, it naturally goes to (\ref{Rum}).

As a derivation, it ensures that consistency
requirements (\ref{Leibniz}) and (\ref{FI}) {\em are} satisfied, with the 
suitable insertion of $\*$-multiplication in the proper locations to ensure
full combinatoric analogy,
\be
[ A\* B, L_x , L_y , L_z  ]_\*= A\* [ B, L_x , L_y , L_z  ]_\*  
+ [ A, L_x , L_y , L_z  ]_\* \* B~,  \label{qLeibnz}
\ee
and  
$$
[ [ L_x ,L_y ,L_z  ,D ]_\* ,E,F,G ]_\*  +  
[ D, [ L_x ,L_y ,L_z  ,E ]_\* ,F,G ]_\* +  
[ D,E, [ L_x ,L_y ,L_z  ,F ]_\* ,G ]_\* 
$$
\be
+[ D,E,F, [ L_x ,L_y ,L_z  ,G ]_\* ]_\*
  =  
[ L_x ,L_y ,L_z  , [ D,E,F,G ]_\* ]_\*  ~.
\ee

The reader might also wish to note from (\ref{QNB}) that, for constant $A$
(independent of phase-space variables), thus $dA/dt=0$, 
\be 
[ A,B,C,D ]_{\*}=0 
\ee
holds identically, in contrast to the 3-argument QNB \cite{nambu}.
Thus, {\em no} debilitating constraint among the arguments $B,C,D$ is 
imposed;  the inconsistency identified in ref \cite{nambu} is a feature of
odd-argument QNBs and does not restrict the even-argument 
QNBs of phase space considered here.

As indicated, in the generic case, the QNB (which provide the 
correct quantization rule for all the systems considered here)  
need not satisfy the 
Leibniz property and FI for consistency, as they are not necessarily 
derivations.  However, they entwine derivations within symmetrized Jordan 
products of invariants: it turns out that the proportionality 
function $V$ of (\ref{tuzsuz}) is not irrelevant\footnote{
If, for some reason related to specific application contexts, 
one still insisted on a bracket structure which {\em is} a derivation, with 
some effort, the relevant structures might be unentwined through infinite 
series solutions of (\ref{q3s}), requiring special attention to 
convergence issues \cite{CQNB}.
}.

For instance, for $S^3$, to quantize (\ref{omunene}) for $N=3$, 
\be
P_{2} \frac{dk}{dt}= \{ k,P_{1},L_{12},P_{2},L_{23},P_{3} \} ~,\label{c3s} 
\ee
note that 
\be
\Bigl  [ k,P_{1},L_{12},P_{2},L_{23},P_{3}\Bigr ] _{\*} 
=3i\hbar^3 \Bigl ( P_2 \* \{ \! \{k,H_{qm} \}\! \} 
+\{\! \{ k,H_{qm}\}\! \} \* P_2 \Bigr ) + {\cal Q} ~,  \label{q3s}
\ee
where ${\cal Q}$ is an $O(\hbar^5)$ rotation, a 
sum of triple commutators of $k$ with 
invariants. Consequently, the proper quantization of 
(\ref{c3s}) is 
\be
\Bigl  [k,P_{1},L_{12},P_{2},L_{23},P_{3}\Bigr ] _{\*} =
3i\hbar^3  \frac{d}{dt} \Bigl  ( P_2 \* k + k\* P_2\Bigr )+ {\cal Q} ~,
\ee
and again reduces to (\ref{c3s}) in the $\hbar\rightarrow 0$ limit,
as ${\cal Q}$ is subdominant in $\hbar$ to the time derivative term.
The right hand side not being an unadorned derivation on $k$, it does not 
impose a Leibniz rule analogous to (\ref{qLeibnz}) on the left hand side, 
and so it fails the FI, at no compromise to its
validity, however.  The $N>3$ case and Chiral Models parallel 
the above through use of fully symmetrized products.

More elaborate isometries of general manifolds in such models are expected to 
yield to analysis similar to what has been illustrated for the prototypes 
considered here. An empirical methodology suggested by a plethora of examples 
\cite{CQNB} argues for, first, manipulating the CNB entries to simplify $V$, 
and to select the entries such that they 
combine into the hamiltonian in the PB resolution of the CNB. 
The corresponding QNB would then be expected to yield entwined structures 
as illustrated above, upon working out its MB resolution, 
through analogous combinatoric operations;  
with the hamiltonian appearing in a MB with $k$ (hence its time derivative), 
now entwined with invariants. 
The resulting intriguing dynamical laws are further discussed in 
T Curtright's talk [these Proceedings, hep-th/0303088].


\begin{theacknowledgments}
We thank Y Nambu, Y Nutku, D Fairlie, and J de Azcarraga for helpful
insights. 
This work was supported in part by the US Department of Energy, 
Division of High Energy Physics, Contract W-31-109-ENG-38, and the NSF Award 
0073390. 
\end{theacknowledgments}

 
\end{document}